**Analysing Factors Affecting the Adoption of Ride-Hailing Services (RHS) in India? A Step-Wise LCCA-MCDM Modeling Approach**

**Eeshan Bhaduri**[0000-0002-7020-0986]
Ranbir and Chitra Gupta School of Infrastructure Design and Management
Indian Institute of Technology Kharagpur, Kharagpur, West Bengal, India, 721302
Email: eeshanbhaduri@iitkgp.ac.in

**Shagufta Pal**
Department of Architecture Town and Regional Planning
Indian Institute of Engineering Science and Technology Shibpur, Howrah, West Bengal, India, 721302
Email: shagufta.besu@gmail.com

**Arkopal Kishore Goswami**[0000-0003-1369-215X]
Ranbir and Chitra Gupta School of Infrastructure Design and Management
Indian Institute of Technology Kharagpur, Kharagpur, West Bengal, India, 721302
Email: akgoswami@iitkgp.ac.in

Word Count: 7025 words + 1 table (250 words per table) = 7275 words

Submitted on 30 July 2022

**ABSTRACT**

The study examines heterogeneity in travel behaviour among ride-hailing services (RHS) users by including attitudes, in order to reinforce conventional user-segmentation approaches. Simultaneously, prioritization of ride-hailing specific attributes was carried out to assess how RHS will operate in a sustainable way. The study initially examines latent heterogeneity in users through a LCCA model, and subsequently prioritizes key RHS-specific attributes for each cluster using and MCDM techniques. Three clusters were identified, based on individuals' attitudes and covariates (socio-demographics, land-use attributes, and travel habits). The largest cluster is termed as *Tech-savvy ride-hailing-ready individuals* (48%), who have higher technological ability, and show maximum acceptance towards RHS. The second largest cluster, termed as *Traditional active-travelling individuals* (28%), display least proclivity towards RHS, probably due to their technological inhibition coupled with their greater attachment to traditional travel modes. Lastly, the *Multimodal PV-loving individuals* (24%) are mostly vehicle-owners but also prefer RHS for occasional trips. The final ranking obtained from the analysis revealed that travel time, reliability, and flexibility (*motivators*); travel cost, and waiting time (*deterrents*) are perceived as the key attributes influencing RHS adoption in the Indian context.

## INTRODUCTION

Ride-hailing services (RHS), since its introduction in March 2009 in San Francisco, have gained in popularity and accessibility. Though RHS operators pledge to encourage shared mobility and replace individual vehicles in order to advance sustainable urban mobility (*1*), these services do attract travellers by offering hedonic benefits equivalent to personal vehicles (*2*). Also, a few topical studies have indicated that it might have unwanted repercussions like induced travel, deadheading, and substitution of more sustainable modes (*3–5*). Prior research on user characteristics indicate that (regular) RHS users are typically young, wealthy, and well-educated, living in dense neighbourhoods of large metropolitan centres (*4, 6, 7*). Nonetheless, majority of the initial research concerning RHS (*7, 8*) relies on descriptive study of potential users. Subsequently studies looked at incorporating trip details and demographics into econometric models (*1, 9–11*). However, despite the fact that integrating subjective factors (attitudinal) offer useful insights into the decision-making process, such aspects have not yet garnered much attention (*12*), up until the last decade. In addition to that, analysis of attitudes gains further importance as differences in ride-hailing usage among various income groups have narrowed over time (*13*) with rising ride-hailing market penetration.

## LITERATURE REVIEW

### Operational factors influencing RHS adoption

Clewlow and Mishra (*14*) highlighted the role of safety, service flexibility and comfort as major motivators for adoption of RHS in the context of USA. At the same time, travel time saving and ability to multitask were found to be least important predictors of RHS usage. In contrary, another study by Wang and Mu (*15*) found factors like less commuting time correlate with higher Uber accessibility. Similarly, travel time, travel cost, and comfort were identified as most significant influencing factors for dynamic ridesharing services. Besides, different land use and travel attributes were found to impact the demand for RHS (*16*). Privacy concerns have been observed to be main deterrents for pooled RHS adoption (*11*). Another study identified sales promotion (travel cost), benefits of booking app (efficiency), and service quality to play main role in loyalty among RHS users (*17*). Hence, the attributes that relate to operational characteristics of RHS can be broadly grouped into two categories- (1) objective (travel time, travel cost, reliability, and flexibility); and (2) subjective (perceived safety, comfort, and efficiency).

### Attitudinal factors influencing mobility preferences

The existing literatures have identified various lifestyle and attitudinal variables (latent constructs) which influences mobility related choices. Lavieri and Bhat (*11*) mentioned four key such variables- privacy sensitivity, tech-savviness, variety-seeking lifestyle and green lifestyle. Alemi et al. (*18*) also proved that attitudes related to technology embracing, variety seeking and pro-environmental policies impact RHS choices vis-à-vis other modes. Other recent studies have established that travellers' attitude towards other modes also play key role in shaping their utilization of RHS (*19*). A few relevant studies in the domain of other emerging transport services (e.g., autonomous vehicles) were also considered to explore and compare latent constructs (*20, 21*). Therefore, the present study employs two categories of mobility-related attitudinal constructs- (1) attitudes towards travel alternatives (personal vehicle and public transit); and (2) lifestyle specific attitudes (environmental awareness, technology adoption, variety seeking, and subjective norms).

Latent-class cluster analysis (LCCA) was employed to find three unobserved classes within a sample of California ride-hailing customers (*9*). In 2018, a follow-up LCCA study in California identified three classes with regard to modal implications (*22*): Personal car augmenters (56%); substituters (23% of the sample); and multimodal augmenters (21%). In a different study, Lee et al. (2022) (*23*) used LCCA to observe the impact of using RHS on other travel modes in Phoenix, Atlanta, and Austin, USA. Although insightful, the earlier researchers place more emphasis on the distinct ride-hailing impacts. As such, there is a lack of clarity in the existing research on the three aspects: (1) *who* are embracing this emerging mobility service and (2) to *what* extent, and finally (3) *how* will RHS operate in a sustainable way.

**STUDY OBJECTIVES & METHODS**
     The review of literature led to identification of the gaps in research, and helped in formulating the study objectives. *Firstly,* there is a paucity in RHS literature that performs latent class segmentation with mobility-related attitudes while using demographics (personal and household), built environment and travel habits. *Secondly,* there is also a need to identify macro-level RHS operational factors as motivators and deterrents. *Thirdly,* there is a dearth of studies exploring the heterogeneity in key attributes for various traveller classes. *Lastly*, there is the need to develop robust aggregated rankings to support long-term policy interventions.

     This study advances prior work by using LCCA modelling with attitudes while incorporating separate sets of demographics, built environment and existing travel characteristics as inactive covariates. Such an effort can not only increase the model's explanatory ability but also yield two advantages- *firstly,* estimation of the LCCA model with general travel and lifestyle related attitudes (e.g., attitude towards other modes, technology adoption) which brings all travellers under purview, irrespective of their current RHS adoption status; *secondly,* developing a prioritization framework for RHS operational attributes for each latent class, so as to overcome the subjectivity of LCCA model and translate the attitudes as policy variables. The analysis involves three steps – (step-1) exploratory factor analysis (EFA), (step-2) Latent class cluster analysis (LCCA), and (step-3) Multi criteria decision making (MCDM) and Meta Ranking. Finally, policy recommendations based on the findings have been elucidated.

**DATA COLLECTION**

**Study context**
The city of Kolkata, West Bengal, India is considered as the study area. It is one of India's million-plus cities, with a high population density (24,306 people per square kilometre) and diverse housing stock. According to an Uber-commissioned survey, Kolkata inhabitants have the largest percentage of commuters (13%) opting for ride-hailing services among four metros in India, which is even higher than Asian countries' average (10%). At the same time, Kolkata residents are the most likely (91%) to buy a car in the next five years, but they are also the most inclined (81%) to abandon that plan if ride-hailing fulfils acceptable service levels (*24*). Because of such unique qualities in terms of RHS use and user preferences, it was found to be an appropriate case study city. The survey region was chosen based on Kolkata's administrative boundaries, which are divided into 15 boroughs, and subsequently into 141 wards.

**Survey design and administration**
Only adult respondents (over the age of 18) were recruited for the survey, which was performed at a household level during three weeks in March 2021. The time period was chosen to capture the post-COVID behaviour and preferences. All main public transportation modes, including bus, suburban rail, and subway (metro rail), had resumed service from November/December 2020.

     Experienced researchers assessed validity of the survey instrument and indicator items, taking into account their relevance and representativeness in light of the desired measures. Furthermore, prior to the main survey, a focus group analysis was conducted to develop a collective consensus, and a pilot research of 50 respondents was undertaken. Finally, a random sampling approach was adopted to achieve substantial representation from varied demographic strata. The entire survey was conducted as a computer-aided (by means of mobile tablets) face-to-face interview, while recording the responses using the Survey Monkey form to minimize human errors and reduce data-entry burden.

     The required sample size value to achieve statistical significance was found to be 384. The study however gathered responses from about 1000 individuals through random sampling and finally estimated 902 data points after cleaning the dataset. Following that, the responses were also double-checked for inconsistency and accuracy of responses across related questions. Thus, un-filled, partially-filled, and erroneous entries were removed. Finally, 839 data points were employed in the modelling process. The survey questionnaire has four major sections – (1) information related to use of RHS in the previous month, (2) attitudinal statements related to use of RHS (collected on a seven-point Likert scale of 1 to 7; 1 being

'strongly disagree' and 7 being *'strongly agree'*), (3) individual demographics, and (4) household socio-demographics.

## Sample characteristics

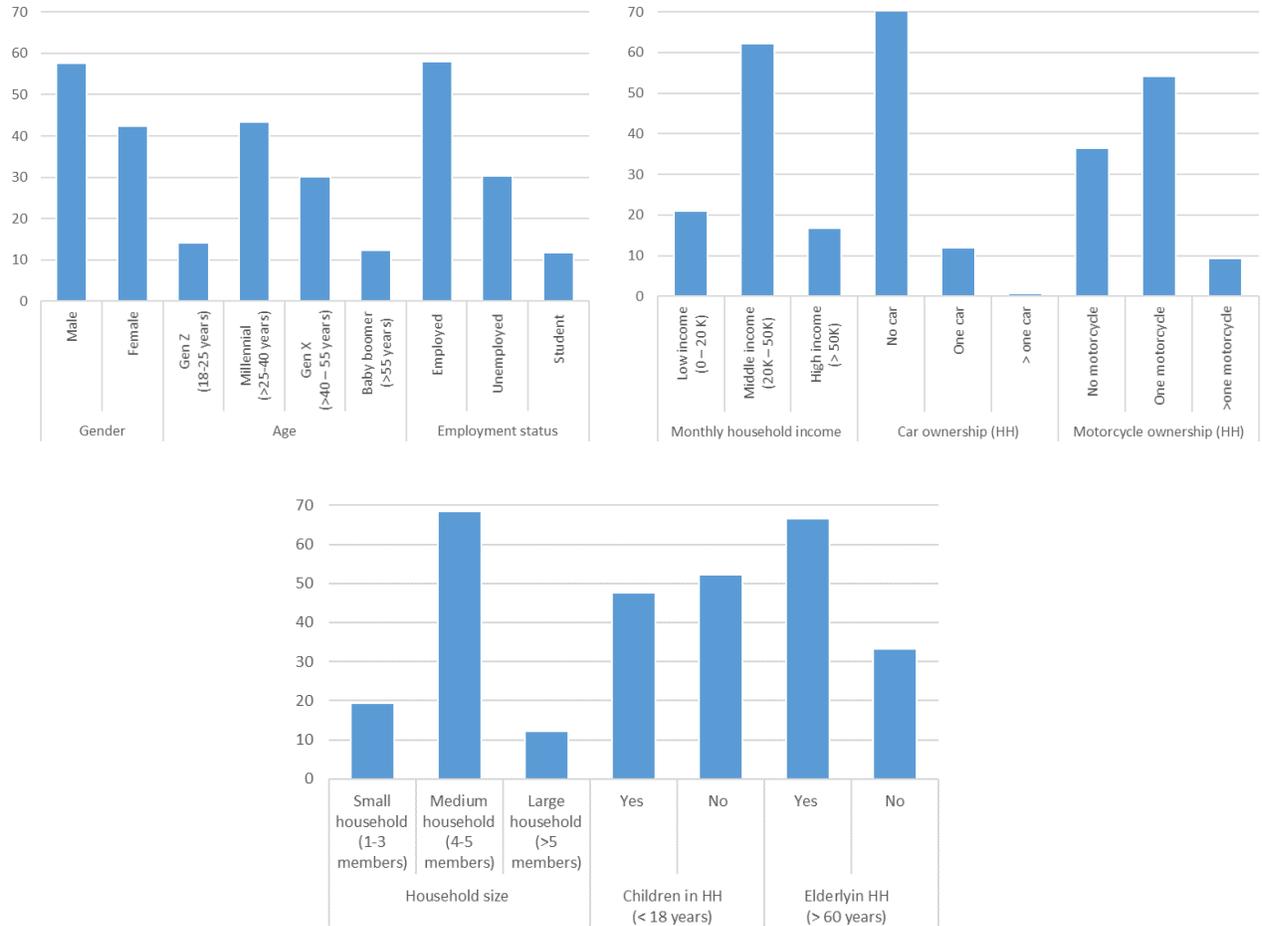

(c)

**Figure 1 Sample characteristics of survey data (% of respondents)**

## Attitudinal indicators

A significant component of the survey exercise incorporated information collection about *mobility specific attitudes* which includes two types of latent constructs- (1) travel alternatives (public transit and personal vehicle); and (2) lifestyle relevant to travel choices. The responses for each of these constructs were collected as level of agreement questions (*25*) in a scale of 1 to 7. Six sets of latent variables were selected based on the conceptual framework (see Figure 2). Each latent variable has been measured with a set of three to four indicators (*26*). The battery of indicator questions for each attitudinal construct was identified from previous behavioural studies, albeit in different domains. Besides, Cronbach's alpha (α) (See Table 1) was calculated to measure the internal consistency of each scale. The final α values are more than the threshold (0.70), indicating sufficient reliability (*27*).

**TABLE 1 Summary of the measured attitudinal statements**

| Latent variable | Code | Question/ Statement | 1 | 2 | 3 | 4 | 5 | 6 | 7 | Mean |
|---|---|---|---|---|---|---|---|---|---|---|
| | | | | | | (%) | | | | |
| *Attitude towards public transit* Cronbach's α = 0.78 | I 1 | I am comfortable to ride on public transit with strangers | 5.0 | 13.1 | 10.7 | 4.5 | 16.0 | 40.5 | 10.1 | 4.75 |
| | I 2 | Riding transit is less stressful than driving on congested highways | 1.2 | 3.6 | 4.8 | 6.0 | 19.4 | 54.1 | 11.0 | 5.45 |
| | I 3 | I feel safe on public transit | 2.7 | 9.2 | 8.5 | 4.5 | 16.6 | 46.5 | 12.0 | 5.11 |
| | I 4 | I like the idea of doing something good for the society when I ride transit | 1.3 | 11.9 | 8.1 | 28.0 | 14.7 | 25.4 | 10.6 | 4.61 |
| *Attitude towards personal vehicle* Cronbach's α = 0.77 | I 5 | I feel I can travel faster in my personal vehicle | 0.2 | 0.6 | 0.7 | 2.1 | 11.7 | 56.4 | 28.2 | 6.07 |
| | I 6 | I feel I am safer in my personal vehicle | 0.1 | 0.6 | 0.7 | 1.7 | 11.6 | 55.4 | 29.9 | 6.10 |
| | I 7 | I feel more convenient in my personal vehicle | 0.4 | 0.1 | 0.1 | 2.1 | 7.5 | 55.8 | 34.0 | 6.20 |
| | I 8 | I feel personal vehicles are status symbols in our society | 10.6 | 22.4 | 4.9 | 16.0 | 24.6 | 14.4 | 7.2 | 3.93 |
| *Tech-savviness* Cronbach's α = 0.76 | I 9 | I frequently use online services (net banking, purchasing products) | 1.1 | 5.4 | 4.1 | 6.2 | 24.7 | 34.8 | 23.8 | 5.48 |
| | I 10 | Getting around is easier than ever with my smartphone | 0.5 | 2.7 | 1.0 | 4.9 | 9.8 | 36.1 | 45.1 | 6.09 |
| | I 11 | I like the idea of using new technologies | 0.4 | 6.0 | 3.1 | 8.6 | 21.6 | 37.1 | 23.4 | 5.50 |
| *Environment-friendly lifestyle* Cronbach's α = 0.80 | I 12 | I would prefer to stay in a walkable neighborhood | 0.6 | 0.7 | 1.9 | 5.5 | 12.8 | 36.4 | 42.2 | 6.07 |
| | I 13 | I would prefer to stay within a 30- minute commute to work | 0.6 | 1.0 | 1.1 | 6.2 | 13.6 | 39.3 | 38.3 | 6.02 |
| | I 14 | When choosing my mode, being environmentally friendly is important to me | 0.4 | 3.3 | 4.9 | 28.0 | 20.3 | 30.8 | 12.4 | 5.06 |
| *Subjective norms* Cronbach's α = 0.76 | I 15 | I might use ride-hailing services if I see someone in my acquaintances (family/ friends) doing so | 1.1 | 11.8 | 4.6 | 31.8 | 16.3 | 25.0 | 9.3 | 4.63 |
| | I 16 | Public opinion will affect my choice of taking ride-hailing services | 0.5 | 19.0 | 6.6 | 24.8 | 20.1 | 19.8 | 9.3 | 4.42 |
| | I 17 | Government policy will influence my choice of taking ride-hailing services | 0.6 | 5.2 | 2.9 | 21.9 | 21.7 | 28.4 | 19.3 | 5.21 |
| *Variety-seeking lifestyle* Cronbach's α = 0.79 | I 18 | I like trying things that are new and different | 3.1 | 12.9 | 6.9 | 16.3 | 24.1 | 26.9 | 9.8 | 4.65 |
| | I 19 | Looking for adventures and taking risks is important to me | 2.6 | 8.8 | 6.7 | 22.4 | 29.1 | 22.5 | 7.9 | 4.66 |
| | I 20 | I love to try new products before anyone else | 3.7 | 19.5 | 10.3 | 21.3 | 17.2 | 18.6 | 9.4 | 4.22 |

1 = Strongly disagree and 7 = Strongly agree for all measures
The mean denoted weighted average of the indicator items

**METHODOLOGY**

In this section, the rationale behind using the stepwise approach have been explained. Moreover, a brief outline of the methods has been included for better comprehension of the current work.

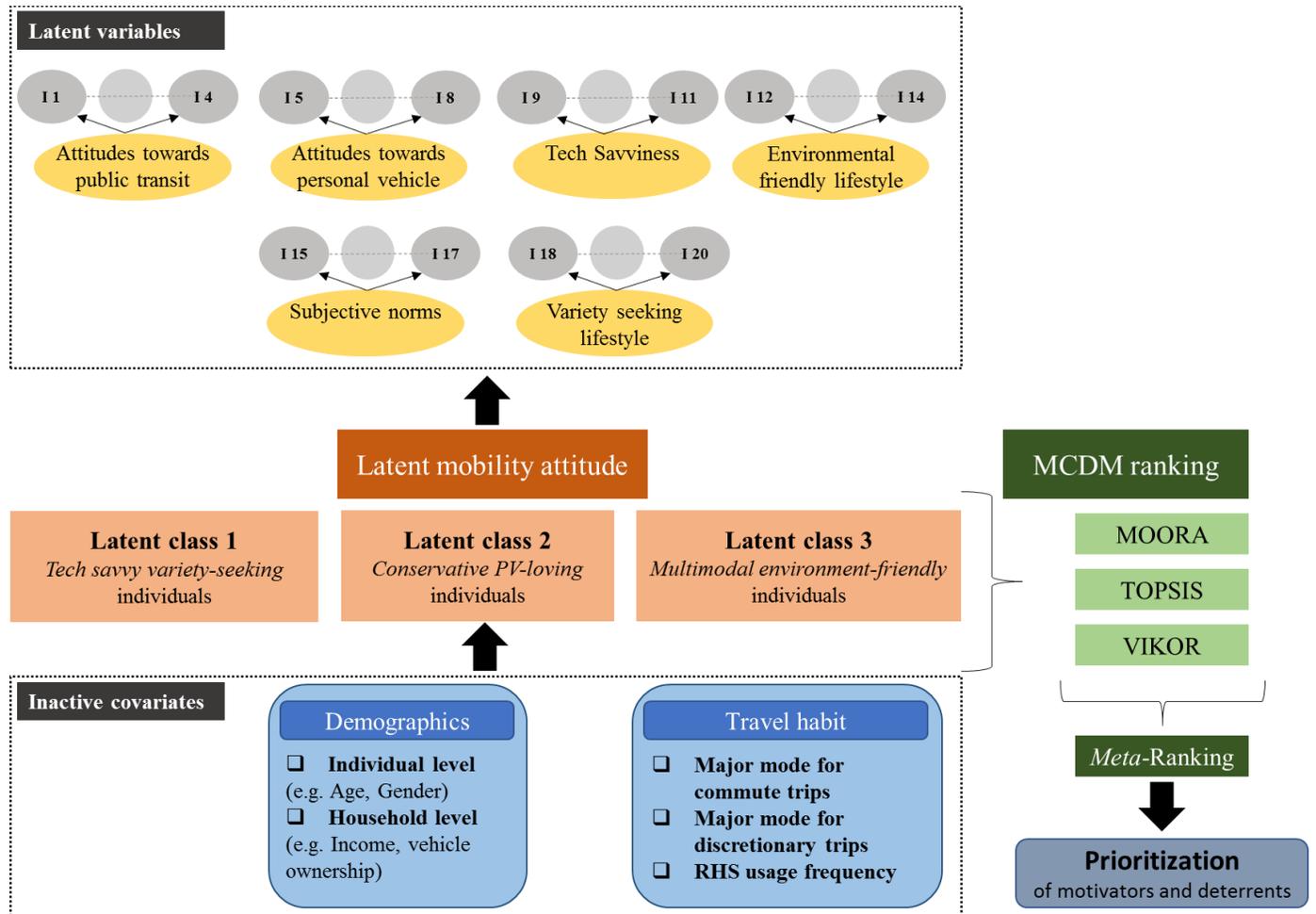

**Figure 2 Structure of two step Latent class cluster analysis (LCCA) with multi criteria decision making (MCDM)**

**Exploratory factor analysis (EFA)**
A total of seventeen indicators related to ride-hailing use have been grouped into six LVs based on an exploratory approach (EFA) utilizing Principal Axis Factoring extraction method with oblimin promax rotation. Using the Kaiser-Meyer-Olkin's (KMO) measure of sample adequacy and the Bartlett's test of sphericity, the data's suitability for EFA was determined. The Bartlett's test of sphericity is less than 0.001, indicating appropriate links between indicators for the EFA, and the KMO value of 0.788 suggests good sample adequacy (*28*). Moreover, the scree plot criterion (*29*) was used to determine the number of factors because the average commonality is lower than 0.6 and the sample size is far over 200 (*30*). As a result, six factors were selected from the factor analysis, accounting for 63.4 percent of the variance. Importantly, only the loadings greater than 0.50 are interpreted (see Figure 3), while the remaining indicators for each factor are marked in grey without any loading. For the posterior LCCA, only these loaded statements are taken into account.

**Latent class cluster analysis (LCCA)**
The factors identified through EFA are used as the indicators of the LCCA model which explains the behavioural heterogeneity of the identified classes. At the same time, the covariates, shown in the lower part of Figure 2, aid in characterizing the different classes. Covariates on demographics, travel habit and built environment-related characteristics were used. Also, they were added only after the model without covariates and adequate fit was identified. The covariates are only included as passive factors to help with cluster identification when they do not improve the model. The interested readers are referred to Molin et al. (*31*) and Alonso et al. (*28*) for further elaboration related to LCCA methodology.

**Multi criteria decision making (MCDM) techniques**
The present study employs three well established MCDM methods: (1) MOORA (Multi-Objective Optimization by Ratio Analysis); (2) TOPSIS (Technique for order of preference by similarity to ideal solution); and (3) VIKOR (VIseKriterijumska Optimizacija I Kompromisno Resenje). MOORA analyses inputs data of each alternative as a variance between cost and benefit criteria (*32*). At the same time, the standard principle of TOPSIS follows that the chosen solution should be closest to the positive ideal solution while being farthest from the negative ideal solution. In this case, positive ideal solutions maximize benefit criteria while minimizing cost criteria (*33*). VIKOR provides solutions for decision problems with conflicting criteria. It possesses certain similarities to TOPSIS as it also introduces a proximity-based ranking index. VIKOR index is calculated based on how important a measure as perceived by respondents with respect to an ideal scenario and worst scenario. Finally, we use Rank Aggreg function (i.e., Meta Ranking function) to derive the combined rank of the attribute by optimizing the ranks which have been earlier obtained from three aforementioned MCDM techniques (*34*).

**RESULTS AND DISCUSSION**
This section presents the estimation results obtained from step-wise LCCA-MCDM modeling framework. It depicts the analysis flow from obtaining the LVs using EFA approach, to using those as indicators for LCCA model, and finally deriving the prioritized attributes (both motivators and deterrents) for each latent class.

**Measurement model**
The six LVs are comparable in terms of number of indicator items, as all of them have three indicators each, except one, which has two indicators (See Figure 3). It should be emphasised that multiple scientific techniques were taken into account to assess the validity of the work. The EFA was also performed in two steps (*35*). As recommended by (*36*), a preliminary EFA was conducted with twenty indicator items to examine the validity and reliability of the constructs in the conceptual model while evaluating the model's suitability by the fit indices. After that, the indicator items with standard factor loadings below the threshold of 0.50 (*37*) were iteratively deleted. The remaining items were then subjected to final EFA, yielding a model with convergent validity supported by extracted average variance, item reliability, and construct reliability.

| Latent variable | Code | Attitudinal indicators | λ | *t*-statistic |
|---|---|---|---|---|
| *Attitude towards public transit* | I 1 | Comfort | 0.758 | 16.416 |
| | I 2 | Stress | 0.542 | 14.832 |
| | I 3 | Safety in public transit | 0.846 | 13.763 |
| | I 4 | Societal responsibility | | |
| *Attitude towards personal vehicle* | I 5 | Time saving | 0.641 | 25.913 |
| | I 6 | Safety in personal vehicle | 0.611 | 23.423 |
| | I 7 | Convenience | 0.652 | 23.542 |
| | I 8 | Status symbolism | | |
| *Tech-savviness* | I 9 | Preference for online services | 0.587 | 29.678 |
| | I 10 | Preference for smartphone | 0.813 | 23.200 |
| | I 11 | Preference for newer technology | 0.649 | 23.186 |
| *Environment-friendly lifestyle* | I 12 | Preference for higher walkability | 0.654 | 17.675 |
| | I 13 | Preference for shorter commute | 0.750 | 19.895 |
| | I 14 | Environmental awareness | | |
| *Subjective norms* | I 15 | Public opinion | 0.535 | 47.549 |
| | I 16 | Acquaintances' opinion | 0.769 | 42.588 |
| | I 17 | Government policy | 0.710 | 49.776 |
| *Variety-seeking lifestyle* | I 18 | Preference for differentness | 0.763 | 22.378 |
| | I 19 | Preference for adventure and risk | 0.575 | 21.869 |
| | I 20 | Preference for newness | 0.765 | 23.564 |

**Figure 3 Standardized parameters of the measurement model**

**Latent class cluster identification**

*Cluster 1* (48% of the sample) - *Tech-savvy ride-hailing-ready individuals:* This cluster, which nearly comprises of almost half of the respondents, has the highest average for three LVs: higher 'tech-savviness', and 'variety-seeking lifestyle' suggest their avant-garde nature, and higher 'subjective norms' indicates their openness to and well informed about the emerging trends (see Figure 4). Further analysis of the demographic covariates shows that cluster 1 has the highest share of younger age groups (Gen Z and Millennials), which could be a precursor to the said behavioural dimensions. At the same time, this cluster also has the lowest values for the other three LVs, indicating their coldness towards environment and reluctance towards using public transit. Such reluctance may be because majority of them are captive public transport users, as reflected by their high usage of bus as major commute travel mode (59%) and lower monthly household income (nearly 53% in low and low-middle income category). Besides, this cluster shows highest affinity towards using ride-hailing services for both commute and discretionary trips.

    *Cluster 2* (28% of the sample) - *Traditional active-travelling individuals:* This cluster includes more than a quarter of the respondents, with slightly greater average values for attitude towards travel modes ('public transit' and 'personal vehicle') as compared to cluster 1. Notwithstanding, it differs from the other two clusters in terms of lowest average for 'tech-savviness', 'variety-seeking', and 'subjective norms', which points out respondents' conservative lifestyle. Here, the term *conservative* needs to be interpreted in light of individuals' affinity for traditional lifestyle including mobility choices. Subsequent analysis of the covariates reveals that they are mostly middle-income users (73%) with lowest car-ownership (10%) but high motorcycle-ownership (76%). Interestingly, this cluster is observed to be the highest user of active travel modes (walk and bicycle) and least users of RHS, which underscores their mentioned lower technological adeptness.

    *Cluster 3* (24% of the sample) - *Multimodal PV-loving individuals:* This cluster which nearly comprises of the remaining quarter of the respondents has the highest average of three latent variables, i.e., attitude towards travel modes ('public transit' and 'personal vehicle'), and 'environment-friendly

lifestyle'. Although while analysing their choices regarding major commute and discretionary trip modes, a higher reliance on personal vehicle (PV), including both car and motorcycle, could be observed. They could be categorized as PV-loving, whose final mode choice depends on attributes like convenience or comfort of personal vehicles, while stated environmental concern may arise out of social desirability bias. Also, this group has been found to use ride-hailing services predominantly for discretionary trips as opposed to commute.

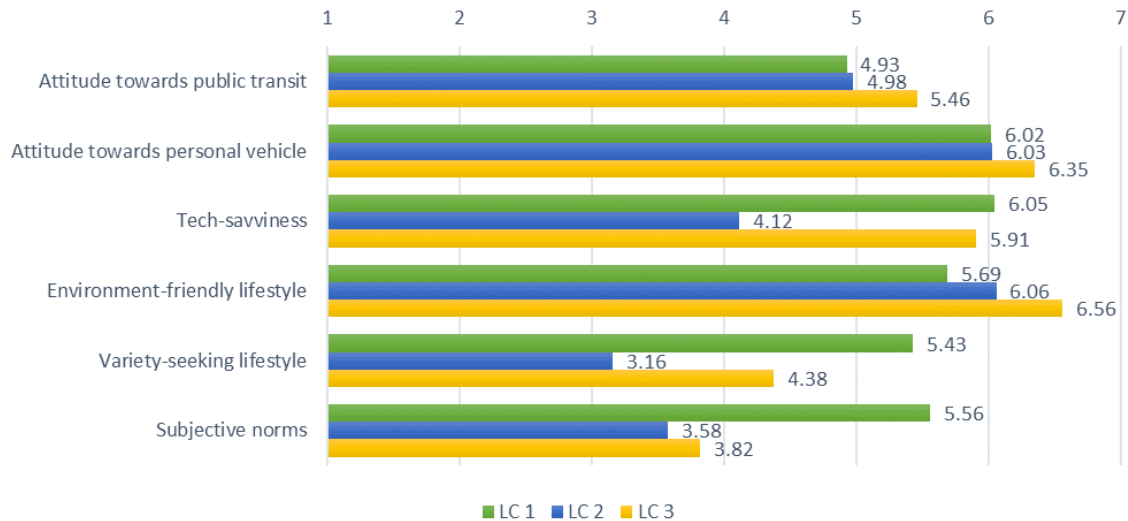

**Figure 4 Average score of six EFA factors (latent variables) for the different clusters**

**Characterization of clusters**
The average EFA scores provide a general idea about cluster identification, but in-depth characterization of the identified clusters is needed with respect to two aspects – (a) sociodemographic characteristics, and (b) existing travel habit. These aspects have been elaborated below.

*Sociodemographic characteristics of the latent clusters*
In the present LCCA model, various socio-demographic attributes (see Figure 5) are analysed as inactive (or passive) covariates, i.e., posterior analysis, after obtaining the class membership function. It has been observed that *Cluster 1* has the highest representation of male respondents (about 61%), which translates into higher variety seeking attitude, as per social psychology literature (*38*). On the other hand, share of female respondents is highest in *Cluster 2*, which is also in line with earlier studies that reveal women having greater propensity towards active travel modes (*39, 40*) while being less variety seeking (*38*).

Intuitively, *Cluster 1* is primarily represented by youth, which can be related to its higher technological affinity (*1, 11*). Interestingly, respondents of *Cluster 1* were found to care least about environment when making travel-related choices. This is not surprising as recent studies on social psychology suggest a decreasing environmental consciousness among younger generation due to over-exposure to materialistic world and over-dependence on technology (*39*). *Cluster 2* includes mostly retired men and women homemakers, who may have lower affinity towards all things digital and also limited travel needs. Such characteristics could also be associated to their greater attachment toward traditional travel alternatives.

In terms of household income groups, *Cluster 1* is associated with low income ones, whereas *Cluster 3* is in complete contrast to it. Further analysis show that *Cluster 3* is more inclined to both car and motorcycle ownership. This may also be explained by the fact that *Cluster 3* is distinct from the other two clusters, with their higher share of households with children (58%) as well as elderly members (73%), which have been acknowledged as motivators for PV fondness (*41, 42*). Besides, car-*dependency* (owners and users) for *Cluster 3* and car-*preference* (non-owners) for *Cluster 2* also stems from the desire of traveling together with their families (for household sizes greater than 3).

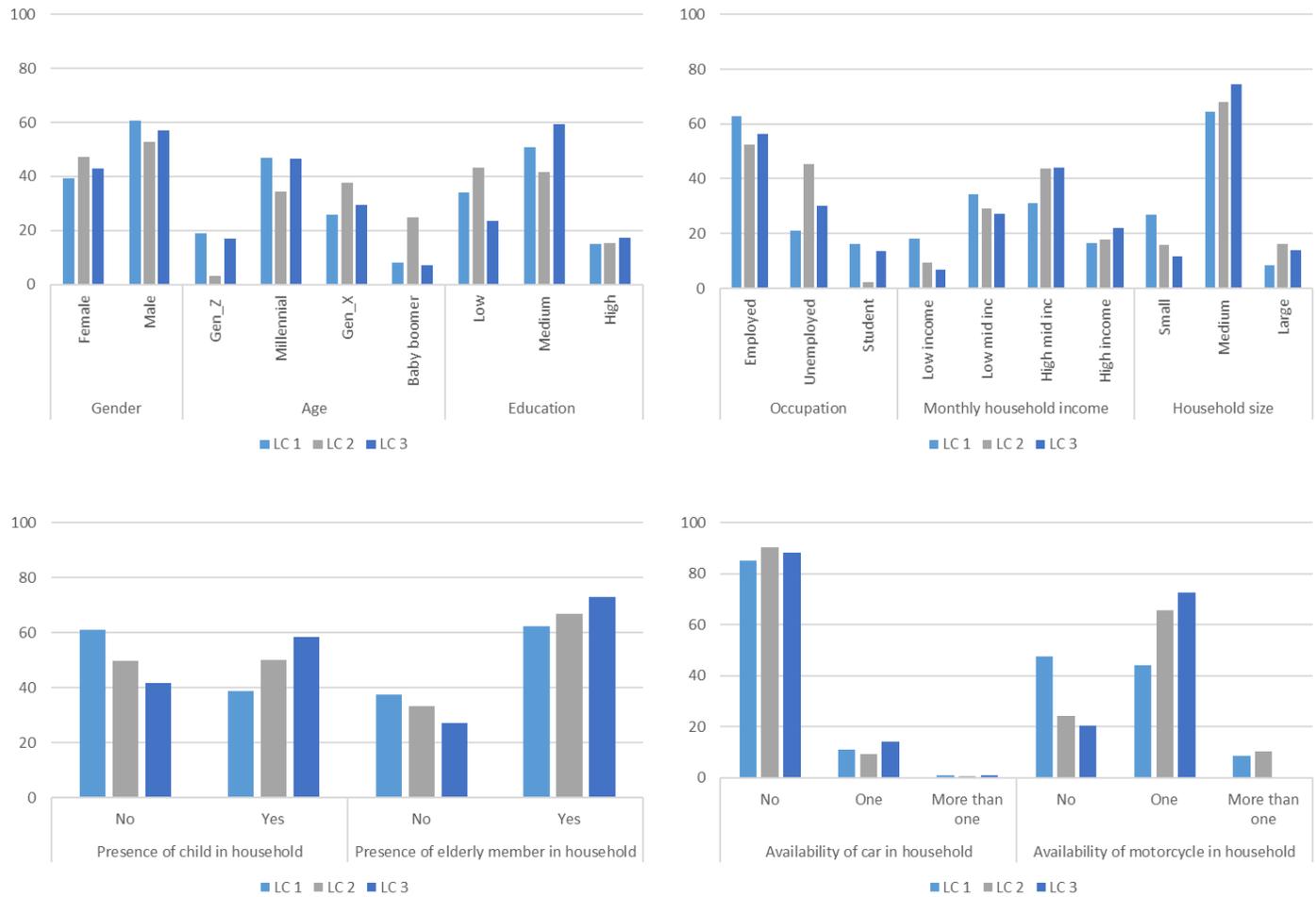

**Figure 5** Profile of the final LCCA model for inactive socio-demographic covariates (% of respondents)

*Existing travel habit of the latent clusters*

*Cluster 1* individuals have the highest share of public transport usage for both mandatory (≅68%) and discretionary trips (≅60%), while *Cluster 3* individuals has the lowest share (≅51% and ≅41% respectively) (Figure 6). Nonetheless, usage of public transit is most recurrent among all three clusters. Interestingly though, the *Cluster 3* individuals are more inclined towards metro rail (subway), especially for discretionary trips, which indicate that higher level of service for metro could attract PV users to public transit. This gains more relevance for *Cluster 3*, which is found to have highest intention of using PVs (car and motorcycle) for both mandatory (≅26%) and discretionary trips (≅25%).

    *Cluster 2* shows maximum usage of active modes (walk and bicycle) for both trip purposes (≅22% and ≅13% respectively). *Cluster 1* respondents choose active modes very rarely, which suggests their participation in time-bounded activities. This is underscored by their highest share in using on-demand vehicles, i.e., taxi and ride-hailing service, which is plausibly because of their ill-experiences in using public transit while being dependent on it.

    The study also analysed heterogeneity among clusters in terms of ride-hailing usage in a month. Around 61% of the respondents in *Cluster 2* have never used it. Understandably, for them the traditional attitude coupled with low technological astuteness have proven to be a hindrance in RHS adoption. When

examining the results further, both *Cluster 1* and *Cluster 3* show resembling trends for the monthly use frequency, which highlights two areas- (1) poor experiences in public transit can push individuals to use RHS (*relevant for cluster 1*); and (2) positive attitude towards personal vehicle can be translated to higher ride-hailing usage (*relevant for cluster 3*). At the same time, it is also observed that *Cluster 1* comprises of more frequent ride-hailing users (≅7%) as compared to *Cluster 3* (≅4%), which presumably could be attributed to higher PV ownership of the latter.

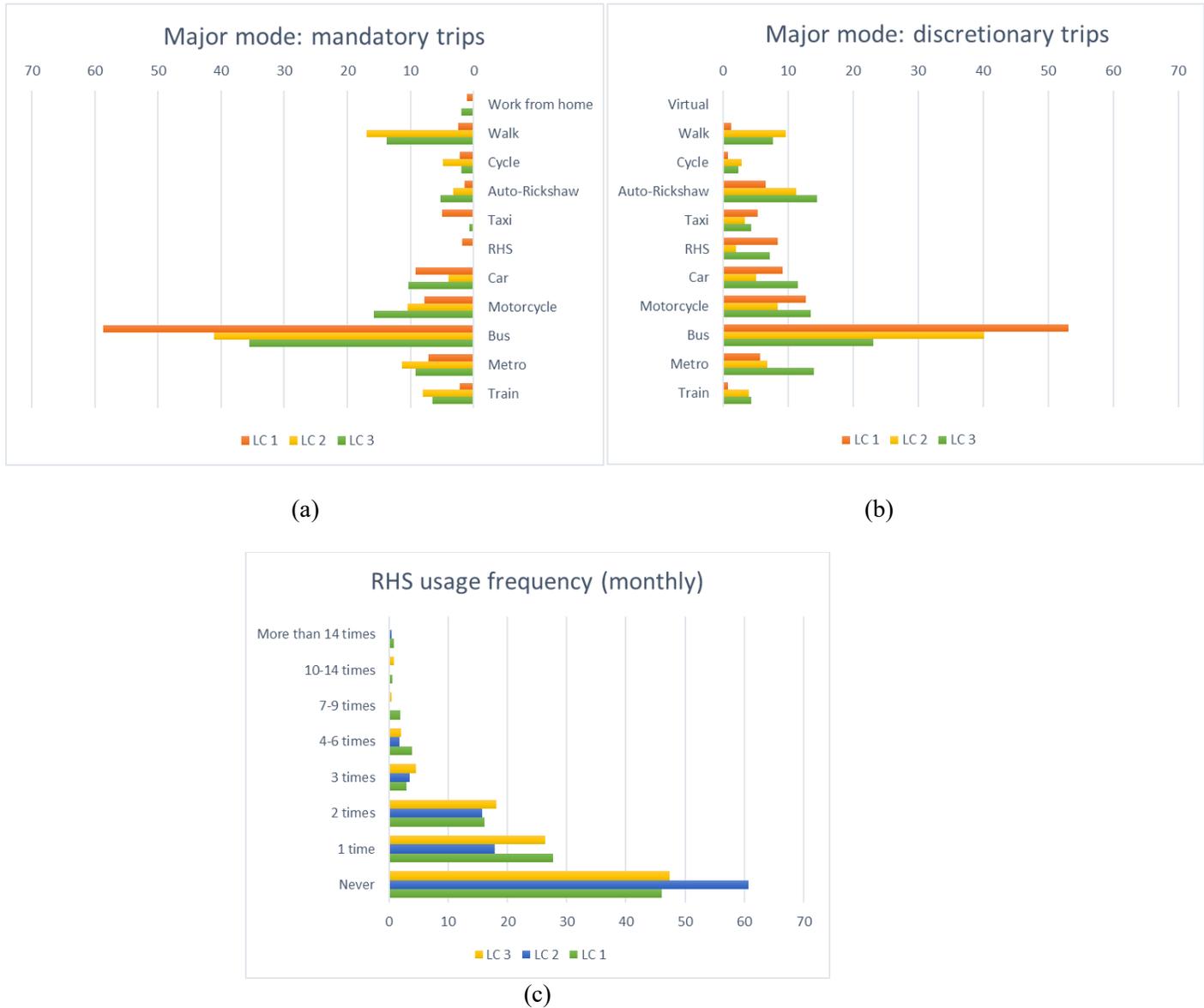

(a)

(b)

(c)

**Figure 6 Profile of the final LCCA model for inactive travel habit covariates (% of respondents)**
# major mode means one which is used most frequently by the respondent

*Built environment attributes of the latent clusters*
The study highlights the impact of built environment on RHS usage from two perspectives, i.e., neighbourhood structure (in terms of population density), and transport infrastructure availability (bus stop density, intersection density, nearest bus stop distance, and road density) (Figure 7). It is observed that population density is not an ideal proxy to be used to understand RHS usage, contrary to suggestion by Lee et al. (2022) (*23*). This might be attributed to the demand-supply disequilibrium created by high population density[1] parameter and lower transport infrastructure availability, as in the case for *Cluster 1* individuals. Expectedly, the impact in terms of higher utilization of motorised modes, especially on-demand services (RHS and Taxi), is more profound for Cluster 1 as compared to Cluster 2. The probable reason being Cluster 1 having a higher percent of regular commuters. Along similar lines, the lowest usage of public transit by Cluster 2 individuals could be related to poor transport infrastructure availability, reflected by greater distance to nearest bus stop.

In fact, it is observed that in context of over-crowded cities (most metro cities in the developing nations), neighbourhood structure density is not the sole dictating factor for travel pattern. However, dense neighbourhood, mostly located near the city cores, simultaneously offer transit and PV-conducive environment by being well connected by public transit and having greater transport infrastructure, in-turn fuelling PV usage. As a result, a specific *Cluster 3* individuals could be ones who concomitantly display PV affection while also being open towards using public transit.

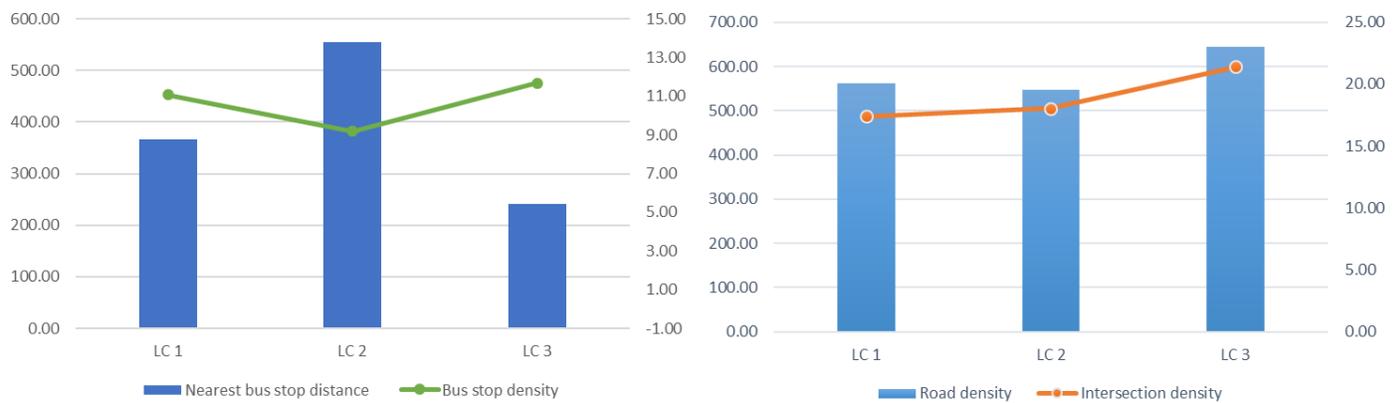

**Figure 7 Profile of the final LCCA model for built environment variables**

**MCDM results**

In this section, the primary motivators and deterrents with respect to using ride-hailing services for the individuals in each of the clusters was analysed. The respondents were presented with a total of twelve factors- six factors in each of the categories (motivators and deterrents). In general, operations related attributes (*for example,* flexibility, travel time, waiting time, travel cost) are observed to weigh more as compared to attributes related to user-perceived benefits (*for example,* safety, health risk, app interface). However, there exists subtle variation among the three clusters, which could assist in policy formulations.

---

[1] Population density (person/sq. Km) for LC 1: 33840; LC2: 26458; and LC 3: 25442 as per Census 2011

| | MOORA | | | TOPSIS | | | VIKOR | | | *Meta Ranking* | | |
|---|---|---|---|---|---|---|---|---|---|---|---|---|
| **Motivating attributes** | LC 1 | LC 2 | LC 3 | LC 1 | LC 2 | LC 3 | LC 1 | LC 2 | LC 3 | LC 1 | LC 2 | LC 3 |
| Flexibility | 1 | 1 | 1 | 1 | 1 | 1 | 1 | 2 | 2 | **1** | **1** | **1** |
| Travel time | 3 | 2 | 2 | 4 | 2 | 2 | 3 | 1 | 1 | **3** | **2** | **2** |
| Reliability | 2 | 4 | 5 | 2 | 4 | 5 | 2 | 4 | 5 | **2** | **3** | **3** |
| Availability | 4 | 5 | 3 | 3 | 5 | 3 | 4 | 5 | 3 | **4** | **5** | **4** |
| Safety | 5 | 3 | 4 | 5 | 3 | 4 | 5 | 3 | 4 | **5** | **4** | **5** |
| Low health risk | 6 | 6 | 6 | 6 | 6 | 6 | 6 | 6 | 6 | **6** | **6** | **6** |

| | MOORA | | | TOPSIS | | | VIKOR | | | *Meta Ranking* | | |
|---|---|---|---|---|---|---|---|---|---|---|---|---|
| **Deterring attributes** | LC 1 | LC 2 | LC 3 | LC 1 | LC 2 | LC 3 | LC 1 | LC 2 | LC 3 | LC 1 | LC 2 | LC 3 |
| Travel cost | 1 | 1 | 1 | 1 | 1 | 1 | 2 | 1 | 1 | **1** | **1** | **1** |
| Waiting time | 2 | 2 | 2 | 2 | 2 | 2 | 1 | 2 | 3 | **2** | **2** | **2** |
| Driver behaviour | 3 | 4 | 4 | 3 | 4 | 6 | 3 | 4 | 4 | **3** | **4** | **4** |
| Online payment issues | 4 | 3 | 6 | 4 | 3 | 4 | 4 | 3 | 5 | **4** | **3** | **5** |
| Customer support | 5 | 6 | 3 | 5 | 5 | 3 | 5 | 6 | 2 | **5** | **6** | **3** |
| App interface | 6 | 5 | 5 | 6 | 6 | 5 | 6 | 5 | 6 | **6** | **5** | **6** |

**Figure 8 Summarized MCDM-based rankings of attributes motivating and deterring RHS adoption**

*Motivators*
The study finds *flexibility* to be the most important motivating factor for adopting ride hailing services across all the three latent clusters (*43*, *44*). Building on Paulssen et al. (2014) (*43*), *flexibility* was explained to the respondents as two distinct abilities- (1) flexibility prior to beginning of a trip, and (2) flexibility during the trip. Importantly, the unanimous appeal of the *flexibility* factor suggests that they could be segregated cluster-wise to obtain better interpretation- (a) the *non-owners* of personal vehicle (*Cluster 1*) could shift towards RHS, which is probably caused by their dissatisfaction with the poor public transit service prevalent in Indian cities (*45*), and (b) the respondents who come from households having more than 3 members, and own personal vehicles, i.e., two-wheelers in case of *Cluster 2*, and four-wheelers for *Cluster 3*, corresponding to a lower vehicle to user ratio, could prefer RHS because of the flexibility they offer.

   *Travel time* and *reliability* are the next two highly rated factors, where the first can be attributed to in-vehicle time, and the latter to schedule reliability. Interestingly, the rankings flip for *Cluster 1*. This underscores the fact that for captive transit riders in Cluster 1, preference for *reliability* stems from the *un*-reliability of existing public transit, which proves to be a greater deterrent than the dis-utility associated with higher travel time (*46*). In addition to this, Cluster 1 is the only cluster which uses RHS for mandatory trips, which has more time-boundedness associated with it. On the other hand, the respondents of *Cluster 2* and *Cluster 3*, where majority of households own at least a two-wheeler, put more emphasis on *travel time*, which is the most plausible reason for their affinity towards PV-ownership. Also, they use it for discretionary trips only, which do not tend to have any strict timeline.

   *Availability* (in adverse conditions) and *safety* (during ride) were found to be associated with low degree of importance. It is worth mentioning that *Cluster 1* and *Cluster 3* show same rankings, whereas *Cluster 2* respondents perceive *safety* to be more important than *availability*. This is likely given that *Cluster 2* is primarily represented by older individuals with a traditional attitude. In contrast to this, the other two clusters comprise of a higher percent of Gen Z and Millennials, who tend to engage more in varied activities even at odd hours (*for example,* late night parties, and un-planned trip with friends) (*47*). It is also worth noticing the lowest importance of health risk benefits of RHS across all three clusters (even after the COVID-19 pandemic). This is likely because they may have developed greater self-reliance (*48*) (*for example,* wearing masks, using personal sanitizers), instead of depending on initiatives from RHS operators.

*Deterrents*
The MCDM analysis results show both, *travel cost*, and *waiting time* to be important deterring factors related to RHS usage across all three clusters. This is in line with recent RHS studies done in the Asian contexts (*49, 50*). However, the bearing of the *travel cost* factor could be profound as it would encourage increased usage of PVs, especially two-wheelers. In other words, *travel cost* deters *Cluster 1* (majorly comprising of low-income individuals) to shift to apparently costlier RHS options, and also fails to attract PV-owners (*Cluster 2 and 3*). The other factor, i.e., *waiting time*, has also been acknowledged as playing a key role for on-demand services (*2, 50*). However, previous studies mostly assume low *waiting time* to be an inherent characteristic of RHS, contrary to the present study.

*Driver behaviour* (includes ride refusals and bad manners) has been perceived to be the third most deterring factor for *Cluster 1*. This might be related to their dependence on RHS in time-constrained situations. Besides, the previous line of reasoning could be extended (higher PV ownership so less impacted by ride refusals) to explain the slightly less importance (fourth-ranked) for *Cluster 2* and *Cluster 3*. *Payment issues* (virtual wallet and online transactions) is observed to be third most deterrent for *Cluster 2*, which could be related to their limited tech-savviness (e.g., internet-based payments) (*51*). Nonetheless, *payment issues* being least important (fifth-ranked) for *Cluster 3* highlight the role of higher income as well as education level. Intuitively, *customer support* has been perceived more important by the frequent RHS user groups, i.e., *Cluster 1* and *Cluster 3*. In line with recent findings (*52*), app interface has been associated with least importance for *Cluster 1* and *Cluster 3*, whereas it is in penultimate position for *Cluster 2*.

## POLICY RECOMMENDATIONS
*Tech-savvy ride-hailing-ready individuals (Cluster 1)*: They are most likely to adopt ride-hailing service given their attitudinal preferences towards newer travel alternatives while having higher technological literacy. Simultaneously, they show higher dissatisfaction towards public transit and reluctance for personal vehicle. As a result, RHS is more likely to substitute public transport which is undesirable from sustainability point-of-view. The plausible solution to this problem could be to provide better multimodal connectivity (for example, Mobility as a Service (MaaS)), while augmenting RHS as last-mile connector. Such option is suitable for this group as they prioritize reliability of rides over lower travel time. Furthermore, MaaS combined with incentivized ride-sharing programmes would facilitate increased acceptance of RHS in lower-income consumer base.

*Traditional active-travelling individuals (Cluster 2)*: They are currently showing least responsiveness towards ride-hailing services, which might be attributed to their technological inhibition as well as being less open to adopt emerging travel means. Not surprisingly, this group highlights online payment issues to be more prominent deterrent relative to the rest groups. Thus, initiatives like booking options through phone call/SMS correspondence could be beneficial. Also, the individuals in this cluster being predominantly older and retired, point towards restricted travel requirements. As such, the ride-safety aspect gains higher importance for this cluster as compared to the rest. Thus, RHS operators need to prioritize safety through ensuring prompt action during emergency and greater accountability of existing safety protocols.

*Multimodal PV-loving individuals (Cluster 3)*: As opposed to the global north (*23, 28, 31*), these individuals are observed to be multimodal and prefer ride-hailing, especially for discretionary trips. This is an ideal scenario to stimulate sustainability by adding dedicated commute shuttle services (for example, *OLA SHUTTLE* service introduced in India) as such an option provides most of the hedonic benefits of PVs, despite being shared service. Moreover, special attention should be given to the fact that this cluster possesses highest share of young families (having children) which is a plausible reason for their proclivity towards PVs. Such growing behavioural inclination and subsequent PV-based travel pattern could be arrested through prioritizing door-to-door service which would cater this group's expectation from RHS, i.e., flexibility, and fast alternative (low travel time).

## CONCLUSIONS

The *Tech-savvy ride-hailing-ready individuals* showed the most acceptance towards RHS. When analysing their priorities regarding RHS, *reliability* emerged as distinctive attribute. The second largest (28%) cluster is the *Traditional active-travelling individuals*, whose members use ride-hailing. Besides, they are more active mode dependent, while their public transit usage is least among three clusters. It can be concluded that stimulation of active modes through planning dense mixed-use neighbourhoods could enhance sustainability. In comparison, *Multimodal PV-loving individuals* (24%) depict strongest affinity for PVs, but their multimodal attitude and sufficient digital acumen can be positively exploited to attract them to shared travel alternatives (e.g., office shuttle service) and even to mass transit. It is worth mentioning that, in accordance with most theories (and applications), our study also makes the implicit assumption that attitudes influence behaviour due to its cross-sectional nature of the data. Therefore, we were unable to disentangle the cause-effect relation between attitudes and behaviour. Hence, we would recommend using panel data in future research to shed further light on these causal processes and to ascertain which causal direction is strongest.

## ACKNOWLEDGMENTS

The authors acknowledge the financial support from the Scheme for Promotion of Academic and Research Collaboration (SPARC), Ministry of Education, Government of India.

## AUTHOR CONTRIBUTIONS

The authors confirm contribution to the paper as follows: study conception and design: E. Bhaduri, A.K. Goswami; data collection: E. Bhaduri, S. Pal; analysis and interpretation of results: E. Bhaduri, S. Pal; draft manuscript preparation: E. Bhaduri, S. Pal, A.K. Goswami. All authors reviewed the results and approved the final version of the manuscript.

# REFERENCES

1.   Alemi, F., G. Circella, S. Handy, and P. Mokhtarian. What Influences Travelers to Use Uber? Exploring the Factors Affecting the Adoption of on-Demand Ride Services in California. *Travel Behaviour and Society*, Vol. 13, No. July 2017, 2018, pp. 88–104. https://doi.org/10.1016/j.tbs.2018.06.002.

2.   Henao, A., and W. E. Marshall. The Impact of Ride-Hailing on Vehicle Miles Traveled. *Transportation*, Vol. 46, No. 6, 2019, pp. 2173–2194. https://doi.org/10.1007/s11116-018-9923-2.

3.   Erhardt, G. D., S. Roy, D. Cooper, B. Sana, M. Chen, and J. Castiglione. Do Transportation Network Companies Decrease or Increase Congestion? *Science Advances*, Vol. 5, No. 5, 2019, p. eaau2670. https://doi.org/10.1126/sciadv.aau2670.

4.   Rayle, L., D. Dai, N. Chan, R. Cervero, and S. Shaheen. Just a Better Taxi? A Survey-Based Comparison of Taxis, Transit, and Ridesourcing Services in San Francisco. *Transport Policy*, Vol. 45, 2016, pp. 168–178. https://doi.org/10.1016/j.tranpol.2015.10.004.

5.   Mahmoudifard, S. M., A. Kermanshah, R. Shabanpour, and A. Mohammadian. Assessing Public Opinions on Uber as a Ridesharing Transportation System: Explanatory Analysis and Results of a Survey in Chicago Area. 2017.

6.   Gehrke, S. R., A. Felix, and T. G. Reardon. Substitution of Ride-Hailing Services for More Sustainable Travel Options in the Greater Boston Region. *Transportation Research Record*, Vol. 2673, No. 1, 2019, pp. 438–446. https://doi.org/10.1177/0361198118821903.

7.   Young, M., and S. Farber. The Who, Why, and When of Uber and Other Ride-Hailing Trips: An Examination of a Large Sample Household Travel Survey. *Transportation Research Part A: Policy and Practice*, Vol. 119, 2019, pp. 383–392. https://doi.org/10.1016/j.tra.2018.11.018.

8.   Clewlow, R. R. Carsharing and Sustainable Travel Behavior: Results from the San Francisco Bay Area. *Transport Policy*, Vol. 51, No. 2016, 2016, pp. 158–164. https://doi.org/10.1016/j.tranpol.2016.01.013.

9.   Alemi, F., G. Circella, P. Mokhtarian, and S. Handy. What Drives the Use of Ridehailing in California? Ordered Probit Models of the Usage Frequency of Uber and Lyft. *Transportation Research Part C: Emerging Technologies*, Vol. 102, 2019, pp. 233–248. https://doi.org/10.1016/j.trc.2018.12.016.

10.  Devaraj, A., G. Ambi Ramakrishnan, G. S. Nair, K. K. Srinivasan, C. R. Bhat, A. R. Pinjari, G. Ramadurai, and R. M. Pendyala. Joint Model of Application-Based Ride Hailing Adoption, Intensity of Use, and Intermediate Public Transport Consideration among Workers in Chennai City. *Transportation Research Record*, Vol. 2674, No. 4, 2020, pp. 152–164. https://doi.org/10.1177/0361198120912237.

11.  Lavieri, P. S., and C. R. Bhat. Investigating Objective and Subjective Factors Influencing the Adoption , Frequency , and Characteristics of Ride-Hailing Trips. *Transportation Research Part C*, Vol. 105, No. May 2018, 2019, pp. 100–125. https://doi.org/10.1016/j.trc.2019.05.037.

12.  Vij, A., and J. L. Walker. How, When and Why Integrated Choice and Latent Variable Models Are Latently Useful. *Transportation Research Part B: Methodological*, Vol. 90, 2016, pp. 192–217. https://doi.org/10.1016/j.trb.2016.04.021.

13.  Jiang, J. More Americans Are Using Ride-Hailing Apps | Pew Research Center. https://www.pewresearch.org/fact-tank/2019/01/04/more-americans-are-using-ride-hailing-apps/. Accessed Jul. 17, 2022.

14.  Clewlow, R. R., and G. S. Mishra. *Disruptive Transportation: The Adoption, Utilization, and Impacts of Ride-Hailing in the United States*. 2017.

15.  Wang, M., and L. Mu. Spatial Disparities of Uber Accessibility: An Exploratory Analysis in Atlanta, USA. *Computers, Environment and Urban Systems*, Vol. 67, 2018, pp. 169–175. https://doi.org/10.1016/J.COMPENVURBSYS.2017.09.003.

16.  Etminani-Ghasrodashti, R., and S. Hamidi. Individuals' Demand for Ride-Hailing Services: Investigating the Combined Effects of Attitudinal Factors, Land Use, and Travel Attributes on Demand for App-Based Taxis in Tehran, Iran. *Sustainability 2019, Vol. 11, Page 5755*, Vol. 11, No.

20, 2019, p. 5755. https://doi.org/10.3390/SU11205755.

17. Nguyen-phuoc, D. Q., D. Ngoc, P. Thi, K. Tran, D. T. Le, and L. W. Johnson. Factors in Fl Uencing Customer ' s Loyalty towards Ride-Hailing Taxi Services – A Case Study of Vietnam. *Transportation Research Part A*, Vol. 134, No. March 2019, 2020, pp. 96–112. https://doi.org/10.1016/j.tra.2020.02.008.

18. Alemi, F., G. Circella, P. Mokhtarian, and S. Handy. Exploring the Latent Constructs behind the Use of Ridehailing in California. *Journal of Choice Modelling*, Vol. 29, No. August, 2018, pp. 47–62. https://doi.org/10.1016/j.jocm.2018.08.003.

19. Acheampong, R. A., A. Siiba, D. K. Okyere, and J. P. Tuffour. Mobility-on-Demand: An Empirical Study of Internet-Based Ride-Hailing Adoption Factors, Travel Characteristics and Mode Substitution Effects. *Transportation Research Part C: Emerging Technologies*, Vol. 115, No. April, 2020, p. 102638. https://doi.org/10.1016/j.trc.2020.102638.

20. Nazari, F., M. Noruzoliaee, and A. (Kouros) Mohammadian. Shared versus Private Mobility: Modeling Public Interest in Autonomous Vehicles Accounting for Latent Attitudes. *Transportation Research Part C: Emerging Technologies*, Vol. 97, No. November, 2018, pp. 456–477. https://doi.org/10.1016/j.trc.2018.11.005.

21. Rahimi, A., G. Azimi, and X. Jin. Examining Human Attitudes toward Shared Mobility Options and Autonomous Vehicles. *Transportation Research Part F: Traffic Psychology and Behaviour*, Vol. 72, 2020, pp. 133–154. https://doi.org/10.1016/j.trf.2020.05.001.

22. Ali Etezady; Patricia Mokhtarian; Giovanni Circella. Investigating the Modal Impacts of Ridehailing and Their Association with Shared Ridehailing.

23. Lee, Y., G. Y. H. Chen, G. Circella, and P. L. Mokhtarian. Substitution or Complementarity? A Latent-Class Cluster Analysis of Ridehailing Impacts on the Use of Other Travel Modes in Three Southern U.S. Cities. *Transportation Research Part D: Transport and Environment*, Vol. 104, No. July 2021, 2022. https://doi.org/10.1016/j.trd.2021.103167.

24. Chin, V., M. Jafar, S. Subudhi, N. Shelomentsev, D. Do, and I. Prawiradinata. *Unlocking Cities- The Impact of Ridesharing across India*. 2018.

25. Likert, R. A Technique for the Measurement of Attitudes. *Archives of Psychology*, Vol. 22, No. 140, 1932, pp. 1–55.

26. Gana, K., and G. Broc. *Structural Equation Modeling with Lavaan*. Wiley-ISTE, 2019.

27. Gliem, J. A., and R. R. Gliem. Calculating, Interpreting, and Reporting Cronbach's Alpha Reliability Coefficient for Likert-Type Scales. *Studies in Inorganic Chemistry*, Vol. 14, No. C, 2003, pp. 349–372. https://doi.org/10.1016/B978-0-444-88933-1.50023-4.

28. Alonso-González, M. J., S. Hoogendoorn-Lanser, N. van Oort, O. Cats, and S. Hoogendoorn. Drivers and Barriers in Adopting Mobility as a Service (MaaS) – A Latent Class Cluster Analysis of Attitudes. *Transportation Research Part A: Policy and Practice*, Vol. 132, No. September 2019, 2020, pp. 378–401. https://doi.org/10.1016/j.tra.2019.11.022.

29. Cattell, R. B. The Scree Test For The Number Of Factors. *https://doi.org/10.1207/s15327906mbr0102_10*, Vol. 1, No. 2, 2010, pp. 245–276. https://doi.org/10.1207/S15327906MBR0102_10.

30. Field, A. Discovering Statistics Using IBM SPSS Statistics. https://books.google.co.in/books?hl=en&lr=&id=c0Wk9IuBmAoC&oi=fnd&pg=PP2&ots=LcEqI F1w0C&sig=qyBgrP_2wPPpSUSwu_eiZqTPGGo&redir_esc=y#v=onepage&q&f=false. Accessed Jul. 30, 2022.

31. Molin, E., P. Mokhtarian, and M. Kroesen. Multimodal Travel Groups and Attitudes: A Latent Class Cluster Analysis of Dutch Travelers. *Transportation Research Part A: Policy and Practice*, Vol. 83, 2016, pp. 14–29. https://doi.org/10.1016/j.tra.2015.11.001.

32. Brauers, W. K. M., and E. K. Zavadskas. Project Management by Multimoora as an Instrument for Transition Economies. *Technological and Economic Development of Economy*, Vol. 16, No. 1, 2010, pp. 5–24. https://doi.org/10.3846/TEDE.2010.01.

33. Lai, Y. J., T. Y. Liu, and C. L. Hwang. TOPSIS for MODM. *European Journal of Operational*

*Research*, Vol. 76, No. 3, 1994, pp. 486–500. https://doi.org/10.1016/0377-2217(94)90282-8.

34. Pihur, V., S. Datta, and S. Datta. RankAggreg, an R Package for Weighted Rank Aggregation. *BMC Bioinformatics 2009 10:1*, Vol. 10, No. 1, 2009, pp. 1–10. https://doi.org/10.1186/1471-2105-10-62.

35. Anderson, J. C., and D. W. Gerbing. Structural Equation Modeling in Practice: A Review and Recommended Two-Step Approach. *Psychological Bulletin*, Vol. 103, No. 3, 1989, p. 411. https://doi.org/10.1037/0033-2909.103.3.411.

36. Jöreskog, K. G., and D. Sörbom. *LISREL 8: User's Reference Guide* . Scientific Software International, Inc., 1996.

37. Hair, J. F., W. C. Black, B. J. Babin, and R. E. Anderson. *Multivariate Data Analysis*. Pearson, 2014.

38. Vianello, M., K. Schnabel, N. Sriram, and B. Nosek. Gender Differences in Implicit and Explicit Personality Traits. *Personality and Individual Differences*, Vol. 55, No. 8, 2013, pp. 994–999. https://doi.org/10.1016/j.paid.2013.08.008.

39. Gifford, R., and A. Nilsson. Personal and Social Factors That Influence Pro-environmental Concern and Behaviour: A Review. *International Journal of Psychology*, Vol. 49, No. 3, 2014, pp. 141–157. https://doi.org/10.1002/IJOP.12034.

40. Gilg, A., S. Barr, and N. Ford. Green Consumption or Sustainable Lifestyles? Identifying the Sustainable Consumer. *Futures*, Vol. 37, No. 6, 2005, pp. 481–504. https://doi.org/10.1016/J.FUTURES.2004.10.016.

41. Van Veldhoven, Z., T. Koninckx, A. Sindayihebura, and J. Vanthienen. Investigating Public Intention to Use Shared Mobility in Belgium through a Survey. *Case Studies on Transport Policy*, Vol. 10, No. 1, 2022, pp. 472–484. https://doi.org/10.1016/J.CSTP.2022.01.008.

42. Md Oakil, A. T., D. Manting, and H. Nijland. Dynamics in Car Ownership: The Role of Entry into Parenthood. *European Journal of Transport and Infrastructure Research*, Vol. 16, No. 4, 2016, pp. 661–673. https://doi.org/10.18757/EJTIR.2016.16.4.3164.

43. Paulssen, M., D. Temme, A. Vij, and J. L. Walker. Values, Attitudes and Travel Behavior: A Hierarchical Latent Variable Mixed Logit Model of Travel Mode Choice. *Transportation*, Vol. 41, No. 4, 2014, pp. 873–888. https://doi.org/10.1007/S11116-013-9504-3/TABLES/4.

44. Vredin Johansson, M., T. Heldt, and P. Johansson. The Effects of Attitudes and Personality Traits on Mode Choice. *Transportation Research Part A: Policy and Practice*, Vol. 40, No. 6, 2006, pp. 507–525. https://doi.org/10.1016/j.tra.2005.09.001.

45. Pucher, J., N. Korattyswaropam, N. Mittal, and N. Ittyerah. Urban Transport Crisis in India. *Transport Policy*, Vol. 12, No. 3, 2005, pp. 185–198. https://doi.org/10.1016/J.TRANPOL.2005.02.008.

46. Majumdar, B. B., M. Jayakumar, P. K. Sahu, and D. Potoglou. Identification of Key Determinants of Travel Satisfaction for Developing Policy Instrument to Improve Quality of Life: An Analysis of Commuting in Delhi. *Transport Policy*, Vol. 110, 2021, pp. 281–292. https://doi.org/10.1016/J.TRANPOL.2021.06.012.

47. Gaffron, P. Urban Transport, Environmental Justice and Human Daily Activity Patterns. *Transport Policy*, Vol. 20, 2012, pp. 114–127. https://doi.org/10.1016/J.TRANPOL.2012.01.011.

48. Zannat, K. E., E. Bhaduri, A. K. Goswami, and C. F. Choudhury. The Tale of Two Countries: Modeling the Effects of COVID-19 on Shopping Behavior in Bangladesh and India. *Transportation Letters*, 2021, pp. 1–13. https://doi.org/10.1080/19427867.2021.1892939.

49. Nguyen-Phuoc, D. Q., D. N. Su, P. T. K. Tran, D. T. T. Le, and L. W. Johnson. Factors Influencing Customer's Loyalty towards Ride-Hailing Taxi Services – A Case Study of Vietnam. *Transportation Research Part A: Policy and Practice*, Vol. 134, No. February, 2020, pp. 96–112. https://doi.org/10.1016/j.tra.2020.02.008.

50. Devaraj, A., G. A. Ramakrishnan, G. S. Nair, K. K. Srinivasan, C. R. Bhat, A. R. Pinjari, G. Ramadurai, and R. M. Pendyala. Joint Model of Application-Based Ride Hailing Adoption, Intensity of Use, and Intermediate Public Transport Consideration among Workers in Chennai City. *Transportation Research Record*, Vol. 2674, No. 4, 2020, pp. 152–164. https://doi.org/10.1177/0361198120912237.

51.    Fu, X. mei. Does Heavy ICT Usage Contribute to the Adoption of Ride-Hailing App? *Travel Behaviour and Society*, Vol. 21, 2020, pp. 101–108. https://doi.org/10.1016/J.TBS.2020.06.005.

52.    Nguyen-Phuoc, D. Q., N. S. Vo, D. N. Su, V. H. Nguyen, and O. Oviedo-Trespalacios. What Makes Passengers Continue Using and Talking Positively about Ride-Hailing Services? The Role of the Booking App and Post-Booking Service Quality. *Transportation Research Part A: Policy and Practice*, Vol. 150, 2021, pp. 367–384. https://doi.org/10.1016/J.TRA.2021.06.013.